\documentclass[twocolumn,aps,prl,showpacs]{revtex4}
\usepackage{graphicx}
\begin{document}
\title{Magnetic neutron scattering study of YVO$_3$: Evidence for an orbital Peierls state}
\author{C.~Ulrich$^1$, G.~Khaliullin$^1$, J. Sirker$^2$, M. Reehuis$^3$,
M.~Ohl$^{4,5}$,\\
S. Miyasaka$^6$, Y.~Tokura$^{6,7}$, and B.~Keimer$^1$}
\affiliation{$^1$Max--Planck--Institut f\"ur
Festk\"orperforschung,
70569 Stuttgart, Germany}
\affiliation{$^2$Theoretische Physik I, Universit\"at Dortmund,
44221 Dortmund, Germany}
\affiliation{$^3$Institut f\"ur Physik, EKM, Universit\"at
Augsburg,
86159 Augsburg, Germany}
\affiliation{$^4$Institut Laue--Langevin, 156X, 38042 Grenoble,
France}
\affiliation{$^5$Forschungszentrum J\"ulich GmbH, 52425
J\"ulich, Germany}
\affiliation{$^6$Department of Applied Physics,
University of Tokyo, 113 Tokyo, Japan}
\affiliation{$^7$Correlated
Electron Research Center (CERC), National Institute of Advanced
Industrial Science and Technology (AIST), Tsukuba 305-8562, Japan}

\date{\today}

\begin{abstract}
Neutron spectroscopy has revealed a highly unusual magnetic
structure and dynamics in YVO$_3$, an insulating pseudocubic
perovskite that undergoes a series of temperature induced phase
transitions between states with different spin and orbital
ordering patterns. A good description of the neutron data is
obtained by a theoretical analysis of the spin and orbital
correlations of a quasi-one-dimensional model. This leads to the
tentative identification of one of the phases of YVO$_3$ with the
``orbital Peierls state'', a theoretically proposed many-body
state comprised of orbital singlet bonds.
\end{abstract}


 \maketitle

Materials with valence $d$-electrons exhibit a multitude of
competing many-body states whose theoretical description is still
in its infancy. Recently, much attention has focused on transition
metal oxides with low-lying electronic states (termed
``orbitals'') in which temperature or doping can drive phase
transitions involving marked redistributions of the valence
electron density (``orbital ordering'') \cite{Tok00}. The orbital
ordering temperatures in insulating oxides (such as the widely
studied manganites) are generally high and approach the
temperatures at which these materials become chemically unstable.
Magnetic phase transitions involving unpaired valence electrons
then occur at much lower temperatures. A striking exception to
this general scenario was recently discovered in insulating
YVO$_3$, where a series of temperature-induced magnetization
reversals heralds a spontaneous redistribution of the valence
electron density far below the magnetic ordering temperature
\cite{Ren98,Nog00,Bla01}.

The microscopic origin of this unusual series of phase transitions
has thus far remained elusive. Given that YVO$_3$ is an insulator
with a simple, nearly cubic lattice structure and only two
localized valence electrons per vanadium atom (3$d^2$), the
difficulty of obtaining a microscopic description of its phase
behavior may seem surprising. It is rooted in the large number of
nearly degenerate many-body states accessible to transition metal
oxides with unquenched orbital degrees of freedom, a situation
termed ``orbital frustration''. Clearly, the orbital occupations
of YVO$_3$ cannot be treated as temperature independent parameters
and are hence affected by orbital frustration. However, while pure
spin models with fixed exchange interactions have been extensively
studied, quantitative solutions of models with variable spins and
orbitals have thus far been obtained only in one spatial dimension
(1D) \cite{Pat98}. In this Letter we show that a 1D spin-orbital
model contains much of the physics underlying the phase behavior
of YVO$_3$. The temperature induced charge density rearrangement
is driven by a thermal crossover between two competing states near
a zero-temperature phase transition of the model: a ferromagnet
and a novel collective singlet state, the ``orbital Peierls
state'', that has been theoretically proposed \cite{Pat98} but
hitherto not been observed. In this state, the superexchange
energy of a system of fluctuating exchange bonds is lowered
through spontaneous formation of a dimerized state with
alternating strong and weak bonds, in analogy to the Peierls
instability of a 1D metal.

The neutron diffraction experiments were carried out on the E5
four-circle diffractometer at the BER II reactor of the
Hahn-Meitner Institute in Berlin, Germany, where Cu- or
PG-monochromators selected the neutron wave vectors
$7.07$~\AA$^{-1}$ and $2.66$~\AA$^{-1}$, respectively. The data
were collected with a two--dimensional position sensitive
$^3$He-detector. The inelastic neutron scattering experiments were
carried out at the IN22 triple--axis spectrometer located at the
Institut Laue--Langevin in Grenoble, France, with a
pyrolytic-graphite (PG) monochromator and a PG-analyzer, both
horizontally focusing. The final wave vector was fixed at either
$2.66$~\AA$^{-1}$ or $4.10$~\AA$^{-1}$. This allowed the
determination of the magnon dispersion relations in an energy
range up to 50 meV, with energy resolutions varying between 1 and
5 meV. The samples were untwinned single crystals of YVO$_3$ ($3
\times 3 \times 11$ mm) grown by a floating zone technique
described earlier \cite{Miy00}.

\begin{figure}
\includegraphics[width=0.95\linewidth]{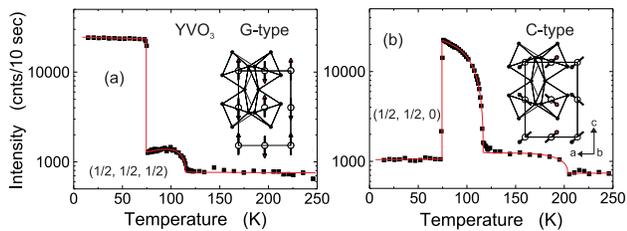}
\caption{\label{fig1} Integrated intensities of (a) the (1/2, 1/2,
1/2) and (b) the (1/2, 1/2, 0) Bragg reflection of YVO$_3$ as a
function of temperature. Both reflections are structurally
allowed, but most of their intensity is of magnetic origin.
Pictorial representations of the magnetic structures are given as
insets (together with the octahedral rotation pattern
corresponding to the orthorhombic cell). A structural phase
transition at 200 K leads to a weak modification of the structural
component of the (1/2, 1/2, 0) reflection
\protect\cite{Nog00,Bla01}.}
\end{figure}

The room temperature crystal structure of YVO$_3$ is described by
the orthorhombic space group $Pbnm$ \cite{Bla01,Zub76} and lattice
parameters $a = 5.2821(9)$~\AA, $b = 5.6144(8)$~\AA, and $c =
7.5283(11)$~\AA. For simplicity we will follow common practice and
index the reciprocal lattice in terms of a pseudocubic subcell
with (almost equal) lattice parameters $a/\sqrt{2}$, $b/\sqrt{2}$,
and $c/2$. YVO$_3$ exhibits two magnetic phases. Between 1.5 K and
$\rm T_{N1}= 75$ K, we observe magnetic Bragg reflections of the
type $(h/2, k/2, l/2)$ with $h$, $k$, $l$ odd (Fig. 1a). The
magnetic structure is thus of G-type, that is, antiferromagnetic
in all three directions of the pseudocubic cell (inset in Fig.
1a). A full refinement of the magnetic structure ($R(F) = 0.034$)
yields an ordered moment of $1.72(5) \mu_{\rm B}$ oriented along
$c$. The magnetic diffraction pattern changes at $\rm T_{N1}$, and
for $\rm T_{N1} < T < T_{N2}=$ 116 K reflections of the type
$(h/2, k/2, l)$, with $h$, $k$ odd and $l=0$ or even, are dominant
(Fig. 1b). In this temperature range, the magnetic structure is
predominantly of C-type, that is, antiferromagnetic in the
$ab$-plane and ferromagnetic along $c$. These observations are
consistent with previous work on powder samples \cite{Zub76}.

However, our single-crystal data now show that the high
temperature phase is much more complex than previously assumed.
For instance, magnetic Bragg reflections of G-type persist in this
phase with diminished intensity (Fig. 1a), indicating that the
spin structure is noncollinear. A refinement of the magnetic
diffraction pattern ($R(F) = 0.093$) yields a G-type component of
$0.30(4) \mu_{\rm B}$ along $c$, and C-type components of $0.49(3)
\mu_{\rm B}$ and $0.89(2) \mu_{\rm B}$ along $a$ and $b$,
respectively. The moments are thus canted by $\theta = 16.5 (1.8)
^\circ$ out of the $ab$ plane, and the total ordered moment is
$1.05(2) \mu_{\rm B}$, much smaller than both the free-ion moment
of $2\mu_{\rm B}$ and the ordered moment of the low temperature
phase at $\rm T_{N1}$. The large reduction of the order parameter
suggests strong quantum fluctuations in the high temperature
phase.

\begin{figure}
\includegraphics[width=0.5\linewidth]{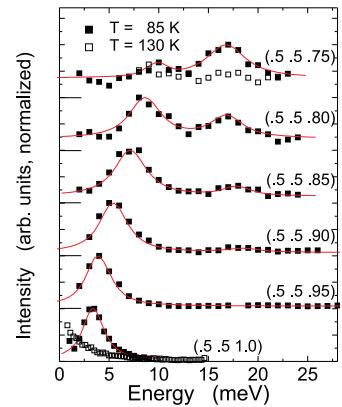}
\caption{\label{fig2} Solid symbols: Inelastic neutron scattering
profiles in the high temperature phase of YVO$_3$, showing
acoustic and optical magnon branches. (Note that $l=1$ is
equivalent to $l=0$.)  The lines are the results of fits to a
magnon cross section convoluted with the spectrometer resolution
function \protect\cite{restrax}. Open symbols: Data taken in the
paramagnetic state.}
\end{figure}

The unusual nature of the high temperature phase is underscored by
its magnetic dynamics. Representative inelastic neutron scattering
data are shown in Fig. 2, and a synopsis of the magnon dispersions
extracted from fits to these data is given in Fig. 3. The magnons
in the low temperature phase follow a simple anisotropic
Heisenberg model with exchange parameters $J_{c} = 5.7 \pm
0.3$~meV and $J_{ab} = 5.7 \pm 0.3$~meV (Fig. 3a). The zone-center
gaps can be attributed to a single-ion anisotropy of the form
$-K_1 S_{iz}^2 + K_1' (S_{ix}^2-S_{iy}^2)$ with $K_1 = 0.33 \pm
0.08$ meV and $K_1' = 0.18 \pm 0.05$ meV. Up to the first order
transition at $\rm T_{N1}$, the G-type magnetic correlations are
weakly temperature dependent (Fig. 1a). In the high temperature
phase, low energy magnons are observed at both $Q=(1/2, 1/2, 1/2)$
and $(1/2, 1/2, 0)$, as expected based on the canted spin
structure (Fig. 1b). The magnon gaps at these points are identical
and substantially smaller than the zone-center gap of the G-phase
immediately below $\rm T_{N1}$, ruling out any two-phase
coexistence scenario.

The spectrum of the high temperature phase (Figs. 2 and 3b)
exhibits several unexpected features. First, the magnon band width
along the ferromagnetic $c$-axis is larger than that in the
antiferromagnetic $ab$-plane. This violates the standard
Goodenough-Kanamori rules according to which ferromagnetic
superexchange interactions are generally substantially weaker than
antiferromagnetic interactions. Further, the spectrum is split
into optical and acoustic magnons with a gap of 5 meV between both
branches. The optical/acoustic splitting, which constitutes a
large fraction of the total magnon band width, can only be
described by assuming two different ferromagnetic exchange bonds
along the $c$-axis, so that the spin Hamiltonian becomes

\begin{eqnarray}
\lefteqn H \;\; &=& \sum_{\langle i,j\rangle \in ab} J_{ab}(\vec
S_i\cdot\vec S_j) - \sum_i K_1 S_{ix}^2 -  \sum_{\langle
i,j\rangle \in c} [J_c (1\pm\delta) \nonumber \\ && (\vec
S_i\cdot\vec S_j) - K_2 S_{ix}S_{jx} \pm d
(S_{ix}S_{jz}-S_{iz}S_{jx})]
\end{eqnarray}

\noindent where $S_x \parallel c$, $K_1$ and $K_2$ are the
single-ion and symmetric superexchange anisotropies, and the
Dzyaloshinskii-Moriya vector $\vec{d}$ is along the octahedral
tilt axis which is staggered along $c$. An excellent fit of the
spectrum can be obtained by  $J_{ab} = 2.6 \pm 0.2$ meV,
$J_c=3.1\pm0.2$ meV, $\delta=0.35$, $K_1=0.90 \pm 0.1$ meV,
$K_2=0.97 \pm 0.1$ meV, and $d=1.15\pm 0.1$ meV (Fig. 3b). The
canting angle extracted from Eq. 1, $\theta= \frac12 \tan^{-1}
[2d/(2J_c-K_1-K_2)] = 14 \pm 1^\circ$, also agrees well with the
observed value.

\begin{figure}
\includegraphics[width=0.95\linewidth]{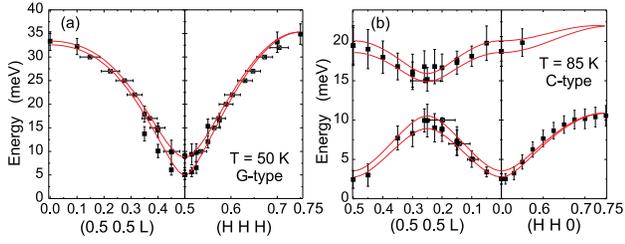}
\caption{\label{fig3}Magnon dispersion relations in the (a) low
and (b) high temperature magnetic phases. The lines result from
diagonalizations of anisotropic Heisenberg Hamiltonians in linear
spin wave theory, as discussed in the text. The effective line
widths, indicated by the bars, can be attributed to the
superposition of two unresolved magnon branches.}
\end{figure}

Because of the crystallographically indistinguishable V-V
distances and bond angles along the $c$-axis \cite{Bla01}, our
observation of strongly alternating exchange bonds in this
direction must be regarded as a signature of highly unusual
orbital correlations. We now proceed to outline a quantum
many-body description of these correlations and their influence on
the spin dynamics; details will be published elsewhere. According
to electronic structure calculations \cite{Saw96,Miz96}, the
vanadium $t_{2g}$-levels are split into a lower-lying singlet of
$xy$-symmetry and a higher-lying doublet spanned by the $xz$- and
$yz$-orbitals. A structural transition associated with a
contraction of the $c$-axis \cite{Bla01} indicates that this basic
hierarchy of electronic states is established around 200 K. The
strong intra-atomic Hund's rule interaction stabilizes a high-spin
($S=1$) state in which the half-occupied $xy$-orbital dominates
the antiferromagnetic in-plane interaction at all temperatures
below 200 K. Below the structural phase transition at 77 K, the
degeneracy of all three $t_{2g}$-levels is split by lattice
distortions, resulting in rigid orbital order with
antiferromagnetic exchange coupling in all directions as observed
experimentally.

According to a recent theory \cite{Kha01}, however, an unusual
electronic state with orbital order involving {\it only} the
$xy$-orbitals is established above 77 K. In this state, the
degeneracy of the $xz$- and $yz$-orbitals, which control the
superexchange interaction along the $c$-axis, is {\it not} lifted
by lattice distortions. Coupling to spin excitations via the
superexchange Hamiltonian broadens the manifold of degenerate
quantum states spanned by these orbitals into a band of overall
width $J=4t^2/U$ (where $t$ is the V-V hopping parameter and $U$
the intra-atomic Coulomb interaction). The band of correlated
spin-orbital fluctuations is {\it one-dimensional}, because only
$c$-axis bonds are involved. (Spin and orbital interchain
interactions are more than an order of magnitude smaller than
$J$.) We now show that this band is unstable against dimerization,
and that this ``orbital Peierls'' instability holds the key to the
unusual magnetic properties we have observed in the C-type phase.

In addition to the overall superexchange energy scale $J$, the
spin-orbital Hamiltonian along the $c$-axis contains two
parameters: the Hund's rule coupling $J_H$ (via the ratio
$\eta=J_H/U$), and the spin orbit (SO) coupling $\Lambda$. To
develop some intuition, we first consider the influence of the
largest parameter, $J$, separately by setting $\eta=\Lambda=0$. In
this limit, the Hamiltonian can be written \cite{Kha01}
\begin{equation}
H = \frac{J}{2} \sum_{<ij>}(\vec{S}_i \cdot \vec{S}_j +1)
(\vec{\tau}_i \cdot \vec{\tau}_j + \frac{1}{4}) \label{Eq1}
\end{equation}
where $<ij>$ denotes a nearest-neighbor pair of V-atoms along $c$,
and $\vec{\tau}$ is a pseudospin-1/2 operator acting in the $xz$-
and $yz$-orbital subspace. The energy of a single exchange bond
described by this Hamiltonian is minimized by formation of an
orbital singlet ($\langle\tau_i \cdot \tau_{i+1}\rangle = -3/4$)
with an accumulation of charge between the V-atoms and
ferromagnetic alignment of the spins. Hence a strong dimerization
in the ground state is expected \cite{She01}.

\begin{figure}
\includegraphics[width=0.95\linewidth]{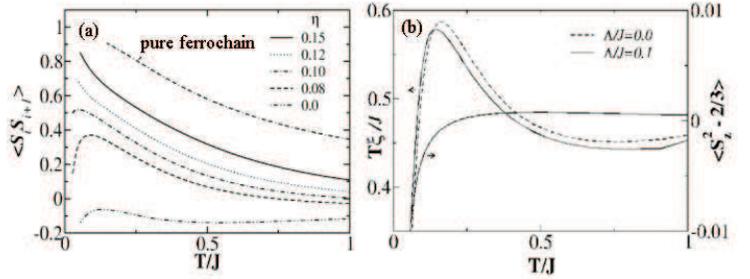}
\caption{\label{fig4} (a) Nearest-neighbor pair spin correlation
function $\langle \vec{S}_i \cdot \vec{S}_{i+1}\rangle$ of the 1D
spin-orbital model described in the text for $\Lambda=0$ and
different values of $\eta = J_H/U$, and of a 1D ferromagnet with
fixed exchange interactions, as a function of temperature. (b)
Correlation length $\xi$ (corrected for the leading $1/T$
divergence for critical excitations) for $\eta=0.15$, and $\Lambda
= 0$ and $\Lambda =0.1 J$ respectively. For the latter parameter
set the spin anisotropy $\langle S_z^2 \rangle - \frac{2}{3}$ is
also shown. }
\end{figure}

In order to elucidate the ground state properties of the full
many-body Hamiltonian, we performed numerical calculations using
the transfer matrix renormalization group (TMRG) method which is
known to yield accurate finite-temperature correlation functions
and thermodynamic properties for a variety of 1D systems
\cite{Pes99,Sir02}. Some results are shown in Fig. 4. The
numerically computed correlator $\langle\vec{\tau}_i \cdot
\vec{\tau}_{i+1}\rangle$ extrapolates to -3/8 per bond at T=0,
thus demonstrating that the ground state indeed spontaneously
dimerizes into orbital singlets on every second bond as suggested
by the simple consideration above (data not shown). While the
exchange interaction between spins within an orbital singlet is
strong and ferromagnetic, nearest-neighbor spins in adjacent
singlets are coupled through a weak, antiferromagnetic exchange
interaction; the spin pair correlator $\langle \vec{S}_i \cdot
\vec{S}_{i+1}\rangle$ is therefore negative (Fig. 4a).

The spontaneous exchange bond dimerization in the ground state of
the Hamiltonian Eq. \ref{Eq1} already hints at an explanation of
the unusual magnetic spectrum observed in the high temperature
phase of YVO$_3$. However, a quantitative description of YVO$_3$
must also include the Hund's rule and SO interactions. The
generalization of Eq. \ref{Eq1} for nonzero Hund's rule coupling
is given in Refs. \cite{Kha01,Sir02}, and the numerically computed
correlator $\langle \vec{S}_i \cdot \vec{S}_{i+1}\rangle$ is
plotted in Fig. 4a for different values of $\eta$. As expected on
general grounds, the ground state for large $\eta$ becomes a
uniform ferromagnet with $\langle\vec{S_i} \cdot
\vec{S}_{i+1}\rangle = 1$. The transition between dimerized and
ferromagnetic states occurs at $\eta_c \sim 0.11$, remarkably
close to the realistic value $\sim$ 0.12 \cite{Miz96} for vanadium
oxides.

Due to the proximity to the zero-temperature phase transition, the
spin and orbital correlations at elevated temperatures exhibit an
intriguing evolution. First, a pronounced maximum of the spin
correlator $\langle \vec{S}_i \cdot \vec{S}_{i+1}\rangle$ as a
function of temperature is apparent for the case of $\eta \lesssim
\eta_c$. The origin of the initial dramatic increase upon heating
lies in the weak exchange bonds between spins in adjacent orbital
dimers. These bonds are highly susceptible to thermal
fluctuations, and the associated increase in entropy stabilizes
the orbital Peierls state. In fact, upon increasing the
temperature this state quickly becomes the leading instability
even for $\eta > \eta_c$, and the numerically computed dimer
correlation length begins to exceed those of other possible ground
states (Fig. 4b). With $J \sim J_{ab}/0.16 \sim 40$ meV extracted
from the low temperature magnon dispersions following Ref.
\cite{Kha01}, the temperature range in which the dimerized phase
is the leading instability coincides approximately with the high
temperature phase of YVO$_3$. This indicates a microscopic
mechanism for the exchange bond dimerization experimentally
observed in this phase. Fig. 4a also shows that the spin
correlations of the spin-orbital chain are frustrated and hence
weaker than those of an orbitally nondegenerate ferromagnetic
chain. The suppression of the spin correlations due to orbital
fluctuations provides a natural explanation for the anomalously
small ordered moment observed in the high temperature phase.

Next, we address the intra-atomic SO coupling $H_{SO} = -2 \Lambda
S_{iz} \tau_{iz}$ where $S_z \parallel c$. TMRG calculations in
which this interaction is explicitly included (Fig. 4b) show that
the interplay between spin-orbit and superexchange interactions
gives rise to an easy-plane spin anisotropy while the
entropy-driven dimerization is almost unaffected. In the
temperature range of interest, the quantity $\langle S_z^2
\rangle$ drops below $\frac{2}{3}$, its value for an isotropic
paramagnet. The preferred spin direction is thus the $ab$-plane,
as experimentally observed in the high temperature phase of
YVO$_3$. These nonlocal orbital correlations are at the root of
the large superexchange anisotropies of Eq. 1. They counteract the
single-ion anisotropy (which is responsible for the $c$-axis
oriented moments in the low temperature phase) and account for the
noncollinear spin structure in the high temperature phase. The
physical origin of this effect is a compromise between the
tendency of the superexchange interaction to co-align the spins
and anti-align the orbital pseudospins of neighboring V-atoms, and
the tendency of the SO interaction to align spin and pseudospin at
every site. In analogy to a spin-flop transition (with the
$c$-axis oriented orbital pseudospin playing the role of the
magnetic field), a ferromagnetic alignment of $ab$-oriented spins
satisfies the superexchange coupling while allowing the spins to
cant out of the plane to take partial advantage of the SO
coupling.

In summary, all of the unusual features of the high temperature
phase of YVO$_3$ including the large ferromagnetic exchange
coupling along the $c$-axis, the optical-acoustic splitting of the
magnon spectrum, as well as the easy-plane anisotropy, the
anomalously small magnitude and the large canting angle of the
ordered moment are thus explained in a quasi-1D model with
dynamical spin and orbital degrees of freedom. Based on this
analysis a lattice dimerization is expected in the high
temperature phase, but may be difficult to observe because of the
large thermal and quantum fluctuations in the orbital sector and
the weak lattice coupling of the $t_{2g}$ orbitals.

\end{document}